\documentstyle[11pt,newpasp,twoside]{article}

\def\edcomment#1{\iffalse\marginpar{\raggedright\sl#1\/}\else\relax\fi}
\reversemarginpar
\input{tcilatex}

\begin{document}

\title{Concluding Perspective}
\author{Martin J Rees}

\affil{Institute of Astronomy, Madingley Road, Cambridge CB3 OHA, UK}

\begin{abstract}
This contribution to the concluding session at IAU Symposium 205
offered comments on some current controversies, including the
scientific status of the ``multiverse'' concept and on the prospects
for cosmology in the coming decade.
\end{abstract}

\section{Preamble}
The last two years have been memorable for cosmologists. The key 
parameters defining our universe's gross properties -- hitherto tentative 
and uncertain -- have seemingly been pinned down more narrowly. Moreover, 
the favoured  values are concordant with a specific consistent model 
(though in some ways an unexpected one).  To put this rapid advance in 
perspective, we should remind ourselves of how the cosmological discourse 
has evolved during the eight decades since astronomers became aware of the 
extragalactic cosmos.

Models for  homogeneous isotropic universes date from the 1920s and 
1930s, and evidence for an expanding universe from the 1920s;  but not 
until the 1950s  was there any prospect of discriminating among the various 
models. Indeed, there was even then little quantitative data on how closely 
any isotropic homogeneous  model fitted   the actual universe.

 A  cosmology meeting in the 1950s would  have focussed on the 
question: is there evidence for evolution, or is the universe in a steady 
state? Key protagonists on the theoretical side would have included  Bondi, 
Gold and Hoyle.  Ryle would have been arguing that counts of radio sources 
-- objects so powerful that many lay beyond the range of optical telescopes 
-- already offered evidence for cosmic evolution; and Sandage  would   have 
advocated the potential of the Mount Palomar 200 inch telescope for 
extending the Hubble diagram far enough to probe the deceleration. 
Intimations from radio counts that the universe was indeed evolving were 
strengthened after 1963 by the redshift data on quasars.

The  era of physical cosmology of course began in 1965, when the 
discovery of the microwave background brought the early `fireball phase' 
into the realm of empirical science, and the basic physics of the `hot big 
bang' was worked out. (The far earlier contributions by Gamow's associates, 
Alpher and Herman, continued, however, to be under-appreciated). There was 
also substantial theoretical work on anisotropic models, etc.  Throughout 
the 1970s this evidence for the `hot big bang'  firmed up,  as did the data 
on light elements, and their interpretation as primordial relics.

Theoretical advances in the 1980s gave momentum to the study of the 
ultra-early universe, and fostered the `particle physics connection': the 
sociological mix of cosmologists changed. There was intense discussion of 
inflationary models,   non-baryonic matter, and so forth.

 Within the last decade, the  pace has been accelerating. This is because 
of a confluence of developments.

   1. {\it The microwave background fluctuations}: these were first detected by 
COBE in 1992, and  are now being probed with enough sensitivity, and on a 
sufficient range of angular scales,  to provide crucial tests of inflation 
and to discriminate among different models.  Ground-based and balloon 
experiments will next year be supplemented by the all-sky coverage of the 
MAP satellite.

2.  {\it The high-redshift universe}: the Hubble Space Telescope (HST) has 
fulfilled ßits potential; two Keck Telescopes have now been joined by the 
VLT,  Subaru, and  Gemini.  We're used to quasars at very high redshifts. 
But  quasars are rare and  atypical -- we'd really like to know the history 
of matter in general.  One of the most important recent  advances  has been 
the detection of many hundreds of galaxies at redshifts up to (and even 
beyond) 5.  Absorption due to the hundreds of clouds along the line of 
sight to quasars probes the history of  cosmic gas in exquisite detail. 
These have opened up the study of `ordinary' galaxies right back to large 
redshifts, and to epochs when they were newly formed. 
And two recently-launched
new X-ray telescopes, Chandra and XMM/Newton,  will offer higher resolution 
and higher sensitivity for the study of distant galaxies and clusters.

3.  {\it Large scale clustering and dynamics}.   Large-scale surveys such 
as the 2dF and the SDSS are vastly enhancing quantitative data on galaxies 
and the statistics of their clustering. Simultaneously with this progress, 
there have been dramatic advances in computer simulations. These now 
incorporate realistic gas dynamics as well as gravity.

4.  {\it Developments in fundamental physics} are offering  new speculative 
insights, which figure prominently in the ongoing debates about the 
ultra-early universe.

\section{The Cosmological Numbers}

 Traditionally,  cosmology was the quest for a few numbers. The first 
were H, q, and $\Lambda$.  Since 1965 we've  had another: the baryon/photon 
ratio. This is believed to result from a small favouritism for matter over 
antimatter in the early universe -- something that was addressed in the 
context of `grand unified theories' in the 1970s. (Indeed,   baryon 
non-conservation seems a prerequisite for any plausible inflationary model. 
Our entire observable universe, containing at least $10^{79}$ baryons,  could 
not have inflated from something microscopic if baryon numbers were strictly 
conserved.)

In the 1980s non baryonic matter  became almost a natural expectation, 
and $\Omega_b / \Omega_{DM}$  is another fundamental number.

 Another specially  important dimensionless number tells us how smooth 
the universe is. It's measured by
\begin{itemize}
\item{} The  Sachs-Wolfe  fluctuations in the microwave background effects
\item{}  the gravitational binding energy of clusters as a fraction of  their 
rest mass
\item{} or by the square of the typical scale of mass- clustering as a 
fraction 
of the  Hubble scale.
\end{itemize}

It's of course oversimplified to represent this by a single number Q, 
but insofar as one can, its value is pinned down to be $10^{-5}$. (Detailed 
discussions introduce further numbers: the ratio of scalar and tensor 
amplitudes, and quantities such as the `tilt', which measure the deviation 
from a pure scale-independent Harrison-Zeldovich spectrum.)

What's crucial is that Q  is small.  Numbers like $\Omega$ and H  are only 
well-defined insofar as the universe possesses `broad brush' homogeneity -- 
so that our observational horizon encompasses many independent patches each 
big enough to be a fair sample. This wouldn't be so, and the simple 
Friedmann models wouldn't be useful approximations,  if  Q weren't much 
less than unity.   Its smallness is necessary if the universe is to look 
homogeneous.  But  it isn't, strictly speaking, sufficient -- a luminous 
tracer that didn't weigh much could be correlated on much larger scales 
without perturbing the  metric.   Simple fractal models for the luminous 
matter are however ruled out by other observational constraints, such as the 
isotropy of the X-ray background, and of the  radio sources detected in 
deep surveys.

\section{How Firmly-Based are Current Models?} 

 After the first millisecond -- after the quark-hadron transition -- 
conditions are  so firmly within the realm of laboratory tests that there 
are no crucial uncertainties in the microphysics (though we should  leave 
our minds at least ajar to the possibility that some `constants' may 
actually be time-dependent). And according to the standard model everything 
is then fairly uniform -- perturbations are still in the linear regime.

It's easy to make quantitative predictions that pertain to this 
intermediate era, stretching from a millisecond to a million years. And 
we've now got high-quality data to confront them with.  The marvellous COBE 
`black body' pins down the  microwave background spectrum to a part in 
10,000. The theoretical framework for light element  nucleosynthesis goes 
back more than 30 years, but the debate (concurrence or crisis?) now 
focuses on 1 per cent effects in helium spectroscopy, and on traces of 
deuterium at very high redshifts. The case for extrapolating back to a 
millisecond is compelling and battle-tested. Insofar as there's a `standard 
model' in cosmology, this is now surely part of it.

 When the primordial plasma recombined, after half a million years, 
the black body radiation shifted into the infrared, and the universe 
entered, literally,  a  dark age. This lasted until the  first stars lit it 
up again. The basic microphysics remains,  of course, straightforward. But 
once non-linearities develop  and bound systems form, gravity, gas 
dynamics, and  the physics in every volume of Landau and Lifshitz  combine 
to unfold the complexities we see around us and  are part of.   The later 
universe, after the dark age is over, is messy and complex -- difficult for 
the same reason that all environmental sciences are difficult.

A decade from now, when the Next Generation Space Telescope (NGST) 
flies, we may know the main cosmological parameters, and have exact 
simulations of how the dark matter clusters. But  reliable knowledge of how 
stars form, when the intergalactic gas is reheated, and how bright the 
first `pregalaxies' are will still involve parameter-fitting, guided by 
observations. The aim is get a 
consistent model that not only  matches all we know about galaxies at the 
present epoch, but also the increasingly detailed snapshots of what they 
looked  like, and how they were clustered,  at all earlier times.

But don't  be too gloomy about  the messiness of  the `recent' universe. 
There are some `cleaner' tests.  Simulations can reliably predict  the 
present clustering and large-scale distribution of non-dissipative dark 
matter. This can be observationally probed by weak lensing,  large scale 
streaming, and so forth, and checked for consistency with the CMB 
fluctuations, which probe the linear precursors of these structures.

\section{Dark Matter: What, and How Much?}

The nature of the dark matter -- how much there is and what it is -- still 
eludes us. It's embarrassing that 90 percent of the universe remains 
unaccounted for.

This key question may yield to a three-pronged attack:

1. {\it Direct detection}.  Astronomical searches for  `machos' in the Galactic 
Halo have shown that only a minority of the halo dark matter could be in 
this form. Indeed nucleosynthesis arguments (especially the D abundance) 
strongly suggest that most dark matter is non-baryonic.  Several groups 
are developing cryogenic detectors to search for supersymmetric particles and axions.

 2. {\it Progress in particle physics}. Important recent measurements suggest 
that neutrinos nave non-zero masses; this result has crucially important 
implications for physics beyond the standard model. However the inferred 
masses seem too low to be cosmologically important.  If theorists could pin 
down  the properties of supersymmetric particles, the number that survive 
from the big bang  could be calculated just as we now calculate the helium 
and deuterium made in the first three minutes.  Optimists may hope for 
progress on still more exotic options.

3. {\it Simulations of galaxy formation and large-scale structure}.  When and how 
galaxies form, the way they are clustered, and the density profiles within 
individual systems,  depend on what their gravitationally-dominant 
constituent is, and simulations are now  constraining the options, and revealing possibly-severe problems with `standard' collisionless cold dark matter.

\section{New Standard Model}
Several lines of evidence suggest that the gravitating dark matter (which 
is predominantly non-baryonic)  contributes substantially less than $\Omega_{DM} = 1$. 

(i) The baryon fraction in clusters is $0.15-0.2$.  On the other hand, the 
baryon contribution to $\Omega$  is now pinned down by deuterium measurements 
to be around  0.015 $h^2$. If  clusters are a fair sample of the 
universe, then this  is incompatible with a dark matter density high enough 
to make  $\Omega_{DM} =1$.

(ii) The presence of clusters of galaxies with $z=1$ is hard to reconcile 
with  the rapid recent growth of structure that would be expected if 
$\Omega_{DM}$ 
were unity.

(iii)  The Supernova Hubble diagram (even though the case for actual 
acceleration may not be compelling)  seems hard to reconcile with the large 
deceleration implied by an Einstein-de Sitter model.

(iv) The  inferred ages of the oldest stars are only barely consistent with 
an Einstein-de Sitter model, for the favoured choices of Hubble constant.

  If there were no evidence other than (i)-(iv) above, then an open model 
with  $0.2-0.3$ would be tenable. But perhaps the most important development 
during the last year has been the strengthening evidence that  the angular 
scale of the `doppler  peaks'  in the CMB angular fluctuations  favours  a 
flat universe
 -- gratifying to 
theorists. But it is a model where baryons make up 4 percent, dark matter 
about 25 percent, and vacuum energy  (or some non-clustered 
negative-pressure component that accelerates cosmic expansion) makes up the 
balance.

The universe is more complicated than some people hoped.  Is it 
contrived that  the vacuum-energy should have the specific small value that 
leads it to start dominating just at the present epoch?

\section{A History of Lambda}
$\Lambda$ was of course introduced by Einstein in 1917 to permit a static 
unbounded universe. After 1929, the cosmic expansion rendered Einstein's 
motivation irrelevant. However, by that time de Sitter had already proposed 
his expanding $\Lambda$-dominated model. In the 1930s, Eddington and Lemaitre 
proposed that the universe had expanded (under the action of the 
$\Lambda$-related repulsion) from an initial Einstein state. $\Lambda$ fell from 
favour after the 1930s: relativists disliked it as a field `acting on 
everything but acted on by nothing'.  A brief resurgence in the late 1960s 
was triggered by a (now discredited) claim for a pile-up in the redshifts 
of quasars at a value of $z$ slightly below 2. The CMB had already convinced 
most people that a the universe emerged from a dense state, rather than 
from an Einstein static model, but it  could  have gone through a coasting 
or loitering  phase where the expansion almost halted. A large range of 
affine distance would then correspond to a small range of redshifts, 
thereby accounting for a `pile up' at a particular redshift. It was also 
noted that this model offered more opportunity for small-amplitude 
perturbations to grow.

The `modern' interest in $\Lambda$ stems from its interpretation as a 
vacuum energy.    The interest has of course been hugely boosted recently, 
through the claims that the Hubble diagram for Type 1A supernovae indicates 
an acceleration. This leads to the reverse problem: Why is $\Lambda$ about 
120 powers of 10 smaller than its `natural' value, even though the 
effective vacuum density must have been very high in order to drive 
inflation? One solution to this dilemma is to postulate a new field (`quintessence') which has, like vacuum energy, a negative pressure but can decay during cosmic expansion. As has been described at this meeting, it is in priciple possible to distinguish this option from a `traditional' time-independent $\Lambda$.

(If $\Lambda$ is fully resurrected,  it will be a great `coup' for de Sitter. 
His model, dating from the 1920s, not only describes inflation, but would also 
describe future aeons of our cosmos with increasing accuracy. Only for the 
50--odd decades of logarithmic time between the end of inflation and the 
present would it need modification!).

\section{Inflation and the Very Early Universe}

\subsection{Testing specific inflation models}

The gross properties of our universe, and indeed its overall scale,
are determined by physics as surely as the He and D abundances -- it's
just that the conditions at the ultra-early eras when the baryon and
dark matter densities, and Q, were fixed are far beyond anything we
can experiment on.

 The inflation concept is the most important single idea. It suggests why 
the universe is so  large and uniform -- indeed, it suggests why it is 
expanding. It was compellingly attractive when first proposed, and most 
cosmologists (with a few eminent exceptions like Roger Penrose) would bet 
that it is, in some form, part of the grand cosmic scheme.  

       Inflationary models still cannot 'naturally' account for the 
fluctuation amplitude $Q = 10^{-5}$  It's important to be clear about the 
methodology and scientific status of such discussion. I comment  with great 
diffidence, because I'm not an expert here.

  The physics of the ultra-early universe remains conjectural, but one
can constrain it by testing particular variants of inflation. For
instance, definite assumptions about the physics of the inflationary
era have calculable consequences for the fluctuations -- whether
they're gaussian, the ratio of scalar and tensor modes, the tilt, and
so forth -- which can be probed by observing large scale structure
and, even better, by microwave background observations.  Measurements
with the MAP and Planck/Surveyor spacecraft will surely tell us things
about `grand unified' physics that can't be directly inferred from
ordinary-energy experiments.

\subsection{A multiverse: is the concept genuinely a scientific hypothesis?}
Some  theories about `extreme physics', when applied to the ultra-early 
universe,  yield many universes that sprout from separate big bangs into 
disjoint regions of space-time. But we do not yet have good reason to trust 
such theories. However, if superstrings (or some 
other equally comprehensive theory)  were `battle tested' by convincingly 
explained  things we could observe, then  if it predicts multiple universes 
we should take them seriously too, just as we give credence to what our 
current theories predict about quarks inside atoms, or the regions shrouded 
inside black holes.

 The `multiverse' is, of course, a highly speculative concept. 
However, the question `Do other 'universes' exist?' is one for scientists 
-- it isn't just metaphysics. The following chain of reasoning may not be 
absolutely compelling, but  should at least  erase any prejudice that the 
concept is absurd.

We can envisage a succession of `horizons', each taking us further 
than the last from our direct experience:

(i) {\it Galaxies beyond range of present-day telescopes}.  There is a
limit to how deep in space, and how far back in time, our present-day
instruments can probe. Obviously there is nothing fundamental about
this limit -- it is constrained just by technology, and enlarges from
year to year.  We would not demote very distant galaxies from the
realm of science simply because they haven't been seen yet -- many
more will undoubtedly be revealed in the coming decades by projected
telescope arrays in space.

 (ii) {\it Galaxies unobservable -- even in principle -- until a remote cosmic 
future} Even if there were absolutely no technical constraints on the power of 
telescopes,  our observations are still bounded by the `particle horizon'. 
This horizon demarcates the `shell' around us on which the redshift would 
be infinite.   There is nothing special about the galaxies on this horizon, 
any more than there is anything special about the circle that defines the 
horizon when you're in the middle of an ocean.  On the ocean, you can see 
further by climbing up your ship's mast. But our cosmic horizon can't be 
extended unless the universe  changes,  so as to allow light to reach us 
from galaxies that are now beyond it.
     
If the expansion were  decelerating, then the far-future  `horizon' 
would  encompass some galaxies that are now undetectable even in principle. 
The `horizon' only grows perceptibly  over the aeons of cosmic evolution. 
It is, to be sure, a practical impediment if we have to await a cosmic 
change taking billions of years, rather than just a few decades (maybe) of 
technical advance,  before a prediction can be put to the test. But does 
that introduce a difference of principle? Surely it is still meaningful to 
talk about these faraway galaxies, and the  far longer time before they can 
be observed is a merely quantitative difference, not one that changes their 
epistomological status?

(iii) {\it Galaxies that emerged from `our' big bang, but are unobservable in 
principle, ever}.
But   what about galaxies that we can never see, however long we wait? 
These are a feature of (for instance) $\Lambda$-dominated cosmological models 
where the expansion accelerates. There would (as in a decelerating 
universe) be galaxies so far away that no signals from them have yet 
reached us.  But if the cosmic repulsion has  overwhelmed gravity, we are 
now accelerating away from them,  so if their light hasn't yet reached us, 
it never will.  Such galaxies wouldn't become observable however long we 
waited.  But does that make them less `real' than they would be if they 
were destined to become  observable  a trillion years hence?

(iv) {\it Galaxies in disjoint universes} The never-observable
galaxies in (iii) would have emerged from the same homogeneous `big
bang' as us. But suppose that, instead of causally-disjoint regions
emerging from a single big bang (via an episode of inflation) we
envisage separate big bangs.  Are space-times completely disjoint from
ours any less real than parts of what we'd traditionally call our own
universe that never come within our horizon?

This four-step argument (some may call it a `slippery slope') tells 
us, I think, that  other universes, even if they are never observable,  are 
within the remit of science.

     If there are other universes, one question that arises is: How  much 
variety might they display? Which features of  our actual universe are 
contingent rather than necessary?   Many would  agree with  Wilczek that 
the most important question in 21st century physics is: `Are the laws of 
physics unique?'  A `final theory'  might  determine uniquely   particle 
masses and coupling constants, and even numbers like Q and the curvature. 
There would then be no role for `anthropic' arguments in cosmology. On the 
other hand, the (still unknown)  underlying laws  that  apply throughout 
the multiverse could turn out to  be more permissive. Each universe may 
then expand in a distinctive way.  Some of its key properties may then   be 
`accidents',  and there would be no explanation for them other than an 
anthropic one.

\subsection{Bayesian tests of whether our universe is drawn from a 
specific type of ensemble}
Suppose  that  the basic numbers describing our universe are an 
arbitrary outcome of how it cooled down,  and take different values in 
other universes. Their values  in our universe may not be typical of the 
entire multiverse: our universe must have been special, and  probably 
highly atypical,   to permit our existence.  But we would need to  think 
again if the numbers  turned   out to be {\it even more special} than our 
presence requires. Even in our present ignorance of fundamental theory, we 
can use a Bayesian argument to test specific hypotheses.

Consider $\Lambda$. An unduly fierce cosmic repulsion would prevent
galaxies from forming. It has to be below a readily calculable
threshold to allow protogalaxies to pull themselves together before
gravity is overwhelmed by cosmical repulsion.  But we wouldn't expect
it to be too far below that threshold.  
Suppose, for instance, that (contrary to current indications)
$\Lambda$ was thousands of times smaller than it needed to be merely
to ensure that galaxy formation wasn't prevented. This would raise
suspicions that it was indeed zero for some fundamental reason.  (Or
that it had a discrete set of possible values, and all the others were
well about the threshold). By this line of argument we could in
principle find strong evidence against specific hypothesis about a
multiverse: the parameters of our universe should not be too atypical
of the anthropically-allowed subset of universes in the ensemble,
weighted by the (theory-generated) prior probability distribution.

\subsection{A historical parallel}

 The multiverse concept might seem arcane, even by cosmological 
standards,    but it affects how we  weigh the observational evidence in 
some current debates. Our universe doesn't seem to be quite as simple as it 
might have been.  It contains atoms, and dark matter;  as an extra 
complication,  there is some kind of `dark energy' in empty space.    Some 
theorists have a strong  prior preference for the simplest universe and are 
upset by these developments.   It now looks as though a craving for such 
simplicity will be disappointed.

   Perhaps we can draw a parallel with debates that occurred 400 years ago. 
Kepler discovered that planets moved in ellipses, not circles. Galileo was 
upset by this. He thought circles  seemed more beautiful;  and they were 
simpler -- one parameter not two.  Newton later showed, however,  that all 
elliptical orbits could be understood by a single unified theory of 
gravity. Had  Galileo still been alive when `Principia' was published, 
Newton's insight would surely have joyfully  reconciled him to ellipses.

    The parallel is  obvious. A universe with low $\Omega_{DM}$, non-zero  $\Lambda$, 
and so forth may seem ugly and complicated.  But maybe this is our limited 
vision.    Our Earth traces out just one ellipse out of an infinity of 
possibilities,  its orbit being constrained only by the requirement that it 
allows an environment conducive for evolution (not getting too close to the 
Sun, nor too far away).  Likewise, our universe  may be just one of an 
ensemble of all possible universes, constrained only by the requirement 
that it allows our emergence. Maybe we
should go easy with Occam's razor and be wary of arguments that
$\Omega=1$ and $\Lambda=0$  are {\it a priori} more natural and less ad
hoc.

\section{The Agenda 10 Years From Now: A Bifurcated Community?}
If we were to reconvene 10 years from now, what would be the `hot 
topics' on the agenda?  The key numbers specifying our universe and its 
content may by then have been pinned down. Or we may discover that our 
universe is too complicated to fit into the framework. I've heard people 
claim that cosmology will thereafter be less interesting -- that the most 
important issues will be settled, leaving only the secondary drudgery of 
clearing up some details.  I'd like to spend a moment trying to counter 
that view.

  It may turn out, of course, that the new data don't fit anywhere within
the parameter-space that these numbers are derived from.  (I was
tempted to describe this as `pessimistic', but of course some people
may prefer to live in a more complicated and challenging
universe!)  On the other hand, maybe everything will fit the
framework, and we will pin down the contributions to $\Omega$ from
baryons, CDM, dark energy, and the vacuum, along with the amplitude
and tilt of the fluctuations, and so forth.  If that happens, it will
signal a great triumph for cosmology -- we will know the `measure of
our universe' just as, over the last few centuries, we've learnt the
size and shape of our Earth and Sun.

   Our focus will then be redirected towards new challenges, as great as 
the earlier ones. But the character and `sociology' of our subject will 
change: it will bifurcate into two sub-disciplines.   In sociological 
terms, this  bifurcation would be  analogous to what actually happened in 
the field of general relativity 20-30 years ago. The `heroic age' of 
general relativity -- leading to the rigorous understanding of 
gravitational waves, black holes, and  singularities -- occurred in the 1960s 
and early 1970s. Thereafter, the number of active researchers in 
`classical' relativity declined (except maybe in computational aspects of 
the subject): most of the leading researchers shifted either towards 
astrophysically-motivated problems, or towards quantum gravity and 
`fundamental' physics.

What will be the foci of the two branches of cosmology we'll be 
pursuing a decade from now? One will be  `environmental cosmology' -- 
understanding the emergence of structure, stars and galaxies. The other 
will focus on the fundamental  physics of the ultra-early universe 
(pre-inflation, m-branes, multiverses, etc). A  few words about each of 
these:

\subsubsection{Environmental cosmology: long range prospects}
 One continuing challenge will be to explore the emergence of structure. 
This is a tractable problem until the first star (or other collapsed 
system) forms. But the huge dynamic range and uncertain feedback thereafter 
renders the phenomena too complex for any feasible simulation.  Even if 
the clustering of the CDM under gravity could be exactly modelled, along 
with the gas dynamics, then as soon as the first stars form we face major 
uncertainties that will still be a challenge to the petaflop simulations 
being carried out a decade from now.  We will still need the NGST to pin 
down what happened in the earliest stages of galaxy formation.

\subsubsection{ Probing  the Planck era and `beyond'}
 The second challenge would be to firm up the physics of the ultra-early 
universe. Perhaps the most `modest' expectation would be a better 
understanding  of the candidate dark matter particles: if the masses and 
cross-sections of supersymmetric particles were known, it should be 
possible to predict how many survive, and their contribution to $\Omega$, with 
the same confidence as that with which we can compute primordial 
nucleosynthesis. Associated with such progress, we might expect a better 
understanding of how the baryon-antibaryon asymmetry arose, and the 
consequence for $\Omega_b$.

 A somewhat more ambitious goal would be to  pin down the physics of 
inflation. Knowing parameters like Q, the tilt, and the scalar/tensor ratio 
will narrow down the range of options.  The hope must be to make this 
physics  as well established as the physics that prevails after the first 
millisecond.

  Better still,  new insights and unifications will tell us whether or not 
that are new scales of complexity far beyond our present horizon.

\acknowledgements 
I am grateful to many colleagues, and to the earlier
speakers at this meeting, for providing the stimulus for these
comments. I thank the Royal Society for support.
\end{document}